# High temperature susceptibility in electron doped $Ca_{1-x}Y_xMnO_3$: Double Exchange vs Superexchange.

H. Aliaga[1], M. T. Causa[1,*], M. Tovar[1], A. Butera[1], B. Alascio[1], D. Vega[2], G. Leyva[2], G. Polla[2], and P. König[2]

[1] Centro Atómico Bariloche and Instituto Balseiro. Comisión Nacional de Energía Atómica and Universidad Nacional de Cuyo. 8400 San Carlos de Bariloche, Río Negro, Argentina.

[2] Centro Atómico Constituyente. Comisión Nacional de Energía Atómica. 1650 San Martín, Buenos Aires, Argentina.

**Abstract**. We present a study of the magnetic properties of the electron doped manganites $Ca_{1-x}Y_xMnO_3$ (for $0 \leq x \leq 0.25$) in the paramagnetic regime. For the less doped samples ($x \leq 0.1$) the magnetic susceptibility, $\chi(T)$, follows a Curie-Weiss (CW) law only for $T > 450$ K and, below this temperature, $\chi^{-1}(T)$ shows a ferrimagnetic-like curvature. We approached the discussion of these results in terms of a simple mean-field model where double exchange, approximated by a ferromagnetic Heisenberg-like interaction between $Mn^{3+}$ and $Mn^{4+}$ ions, competes with classical superexchange. For higher levels of doping ($x \geq 0.15$), the CW behaviour is observed down to the magnetic ordering temperature ($T_{mo}$) and a better description of $\chi(T)$ was obtained by assuming full delocalization of the $e_g$ electrons. In order to explore the degree of delocalization as a function of T and x, we analyzed the problem through Montecarlo simulations. Within this picture we found that at high T the electrons doped are completely delocalized but, when $T_{mo}$ is approached, they form magnetic polarons of large spin that cause the observed curvature in $\chi^{-1}(T)$ for $x \leq 0.1$.

*Corresponding author. E-mail: causa@cab.cnea.gov.ar

Short Title:
# High temperature susceptibility in electron doped $Ca_{1-x}Y_xMnO_3$

## 1. Introduction

Manganites, responding to the formula $B_{1-x}A_xMnO_3$ (with B = divalent alkaline earth and A = trivalent rare earth), have prompted a burst of research activity in the last decade, as they enclosed not only very rich physics but also interest in technological applications. The dynamics of these systems is mainly governed by the Mn ions, whose average valence changes with x between 4+ and 3+. The $Mn^{4+}$ ions have non-compensated spins in $t_{2g}^3$ configuration that give rise to localized S = 3/2 spins. On the other hand, the $Mn^{3+}$ ions have an extra $e_g$ electron (with s = ½), that couples ferromagnetically (FM) to the $t_{2g}$ spins. The $e_g$ electrons tend to be itinerant and lower their kinetic energy by polarizing with FM character the localized $t_{2g}$ spins. This process is known [1] as double exchange (DE) and in these materials competes with the classical superexchange (SE) interaction between Mn ions.

In $Ca_{1-x}La_xMnO_3$ were the whole series, from x = 0 to x = 1, can be obtained with the perovskite structure, the behaviour in the electron and hole doping regions were found different. The parent compounds are indeed quite different: while $LaMnO_3$, a Jahn-Teller system, displays A-type [2] antiferromagnetism, where ferromagnetic (FM) planes order antiferromagnetically (AFM), the non Jahn-Teller $CaMnO_3$ is a G-type antiferromagnet, where each magnetic moment orders AFM with his nearest neighbours [2]. Most studies have been performed on hole doped compounds where ferromagnetism and colossal magnetoresistance (CMR) were found. In the range of electron doping, transport and magnetization studies on $Ca_{1-x}La_xMnO_3$ showed that a complete FM state is never reached and different models were discussed in order to explain the magnetic properties at low temperatures [3, 4]. These studies showed that in this region of doping new and very interesting phenomena appear as a consequence of the $A^{3+}$ substitution for $Ca^{2+}$. In recent works, Granado et al [5] and Neumeier et al [6] reported high temperature susceptibility ($\chi$) measurements in electron doped $Ca_{1-x}La_xMnO_3$ and found that the Curie-Weiss temperature ($\Theta$) presents a rapid increase for small values of x going from $\Theta = -429$ K for x = 0 to $\Theta = 25$ K for x = 0.05. This behaviour was related with changes in the phonon spectra found in Raman Spectroscopy results [5].

We study here the system $Ca_{1-x}Y_xMnO_3$. As in the case of the $La^{3+}$ substitution the non magnetic character of $Y^{3+}$ ions makes this series an excellent system to study the evolution of the magnetism of the Mn ions, without interference with another magnetic species. In a previous work [7] we showed transport and magnetization measurements on this series in the low T regime. We found that small Y concentration produces an important decrease in the electrical resistivity and an increase of the magnetization although complete saturation was not achieved. For x > 0.15, signatures of charge ordering (CO) were found in both, transport and magnetization properties. In this work we present measurements of $\chi(T)$ in the PM regime for $0 \leq x \leq 0.25$. The analysis that we present, based in the $Ca^{2+}$ substitution for $Y^{3+}$ experiments, can be extended to the other electron doped $CaMnO_3$ manganites were a similar behaviour was observed [5, 6].

## 2. Results

Ceramic polycrystalline samples of $Ca_{1-x}Y_xMnO_3$ were prepared by solid state reaction methods [8]. For $0 \leq x \leq 0.25$ we obtained single phase samples and room temperature x-rays diffractograms were indexed in the orthorhombic *Pnma* structure. These samples are only slightly oxygen deficient (< 1%) and, as discussed in [8] for x = 0.10, the magnetic properties are almost unaffected for this level of non stoichiometry. In figure 1(a) we show the evolution with x of the cell parameters: along the series $b/\sqrt{2}c$ varies from $b/\sqrt{2}c < 1$ (O-phase) for x = 0 to $b/\sqrt{2}c \approx 1$ for x = 0.25. The structural distortion increases with x and can be evaluated by the orthorhombic deformation [9] $D = (1/3) \Sigma_i |a_i - <a>|/<a>$ where $a_i$ are the cell parameter *a*, *c*, and $b/\sqrt{2}$ and $<a> = (abc/\sqrt{2})^{(1/3)}$ (see inset in figure 1(a)). As is described in [8], the tilt angle of the $MnO_6$ octahedron, produced by the shift of apical O away from the *b* axis, varied from 10° for x = 0 to 16° for x = 0.25. The rotation angle around the *b* axis remains approximately constant (11.5°). In spite of the differences between the ionic radii of $Y^{3+}$ and $Ca^{2+}$ ($r_Y < r_{Ca}$) in our compounds, a volume increase is observed along the series, as is seen in figure 1(b). In this figure we compare the x dependence of the cell-volume in our samples [10] with that of other $Ca_{1-x}A_xMnO_3$ (A = La, Pr, and Tb) manganites [11-13]. A linear dependence with a positive slope is found in all cases. In the case of A = Pr, where $r_{Ca} \approx r_{Pr}$, the normalized rate of increase is $\approx$ 14%. This increase is a consequence of the larger ionic radius of $Mn^{3+}$ as compared with that of $Mn^{4+}$. The size of the $A^{3+}$ ions only makes this slope larger or smaller: 16% for La ($r_{Ca} < r_{La}$) and 11% and 9% for $Tb^{3+}$ and $Y^{3+}$, respectively ($r_{Tb}, r_Y < r_{Ca}$).

Decreasing T from 300 K, the structure remains orthorhombic (O-phase) for x = 0 [5]. As a function of x, Vega et al [14] found different behaviour. The compound with x = 0.07 remains orthorhombic (O) in the whole T range. For x = 0.15, coexistence with a second orthorhombic (O´-phase) was present below 140 K and, for x = 0.25, a complete transition from O to O´ was present near room temperature.

In figure 2 we show the values for the electrical conductivity ($\sigma$) for T $\approx$ 300 K, as a function of x, as extracted from the data of [7]. The sample with x = 0 is the most resistive one and we observe a monotonous increase of $\sigma(x)$ up to x $\approx$ 0.15. This behaviour is consistent with the progressive importance of the DE with the electron doping produced by the presence of $Y^{3+}$ and precludes, at least at room temperature, the existence of phase separation that would produce a percolative behaviour at a critical concentration. For x > 0.15, $\sigma(x)$ shows a strong decrease indicating that DE becomes less effective. This behaviour may be related to the increase of the crystalline distortions, shown in figure 1(a) that leads to the appearance of the orthorhombic O´-phase, at T < 250 K, for x $\geq$ 0.15. In this series of compounds, the presence of an orthorhombic O´-phase was associated [14] with charge ordered states where the DE is no longer operative.

The magnetization, measured with a Faraday balance magnetometer for T > 300 K and with a SQUID for T < 300K, was, in all cases, linear with H for T > 140 K. Pure $CaMnO_3$ is a G-type AFM that orders [2] at $T_N$ = 123 K. In the paramagnetic (PM) region, the susceptibility, $\chi(T)$, measured in well oxygenated $CaMnO_3$ samples [15], was fitted to a Curie-Weiss law with a Curie constant, C = 2.25(5) emu-K/mol. Since in this case all the Mn sites are occupied by $Mn^{4+}$ ($3d^3$) ions with spin S = 3/2, we expected a C value of 1.875 emu-K/mol. In order to solve this discrepancy, the effects of the Van Vleck T-independent susceptibility, $\chi_{VV}$, should be analyzed. Lines [16] derived an expression for the case of $Ni^{2+}$ ions (which present the same cubic crystal field structure as $Mn^{4+}$) $\chi_{VV} = 8N\mu_B^2 / \Delta$, where N is the density of magnetic ions and $\Delta$ is the energy gap between ground and first excited levels. Taking $\Delta$ = 24000 $cm^{-1}$, as derived by Müller [17] for $Mn^{4+}$ diluted in the perovskite $SrTiO_3$, we estimate $\chi_{VV} \approx 9\times10^{-5}$ emu/mol. Subtracting this contribution from the raw data of [15], a Curie-Weiss behaviour is still obtained but now with C = 1.91(5) emu-K/mol and Θ = -370 K. The large value of Θ indicates a strong AFM superexchange interaction between $Mn^{4+}$ ions.

In figure 3 we show χ(T) for different values of x. At high temperatures, the samples with x > 0 also follow a CW law, although important deviations were found for T ≤ 450 K, as seen in figure 3(a). For x ≤ 0.10, $χ^{-1}(T)$ shows a negative (convex) curvature that is clearly shown in figure 3(b) with an enlarged T scale. From the $χ^{-1}(T)$ data for T > 450 K we have determined the values for C and Θ plotted as a function of the doping x in figures 4 and 5(a), respectively. Here, we have subtracted a Van Vleck contribution (1-x)$χ_{VV}$. We note that this contribution is always small relative to the CW term (≤ 6% at the highest temperatures) and comparable to the experimental uncertainty. The Curie constant follows approximately a linear dependence with x, as shown in figure 4, where the continuous line corresponds to a mixture of $Mn^{4+}$ and $Mn^{3+}$ ions:

$$C = (1-x) C_4 + x C_3 \qquad (1)$$

with $C_4$ = 1.875 emu-K/mol and $C_3$ = 3 emu-K/mol. Notice that small Y doping causes large changes in Θ (figure 5(a)), indicating a rapid evolution from a strong AFM for x = 0 to a FM compound for x > 0.04. This variation accompanies the dramatic changes in the electrical conductivity.

For the samples with 0.15 < x ≤ 0.25 (see figure 3(c)), the susceptibility continue to increase. The high temperature CW regime, with Θ clearly FM, extends well below 450 K. For x = 0.20 and 0.25, $χ^{-1}(T)$ show even a small positive (concave) curvature characteristic of FM manganites [18], above ≈ 250 K. Below this temperature a CO process installs [7, 14] that frustrate the incipient FM ordering. As a consequence, the magnetization stops to increase and AFM ordering is finally observed with a weak-FM component [7]. The transition from a FM-like to a AFM-like susceptibility is the origin of the well defined peak in χ(T) for x = 0.25 (see figure 3(c)).

We may now look at the magnetic ordering temperatures, $T_{mo}$, obtained from low T magnetization measurements [7] in order to compare with the x dependence of Θ. As shown in figure 5 the behaviour is quite different. For x = 0, $T_N$ is much lower than |Θ| due to the competition between first and second neighbour interactions [19]. For low doping, $T_{mo}$ is slightly depressed with x while the AFM-like Θ tends rapidly to zero, becomes FM-like, and reaches Θ = 180 K for x = 0.25. Around x ≈ 0.05, $T_{mo}$ presents a minimum value of ≈ 100 K. Above this concentration $T_{mo}(x)$ becomes almost constant (≈ 110 K).

## 3. Mean Field model

In order to describe the PM behaviour in the simplest terms we may consider the system as a randomly distributed mixture of localized $Mn^{3+}$ and $Mn^{4+}$ ions with concentrations (1-x) and x respectively. Then, in a mean field picture we have

$$M_4 = (C_4/T) [H + (\gamma_{44} + \gamma'_{44})(1-x)M_4 + (\gamma_{43} + \gamma'_{43})xM_3] \qquad (2)$$

$$M_3 = (C_3/T)[ H + (\gamma_{34} + \gamma'_{34})(1-x)M_4 + (\gamma_{33} + \gamma'_{33})xM_3] \qquad (3)$$

where subindexes 4 and 3 indicate the magnetization of $Mn^{4+}$ and $Mn^{3+}$, respectively and $\gamma_{ij}$ are the parameters describing the exchange coupling between $M_i$ and $M_j$. Primed parameters indicate second neighbours interaction. The total magnetization is given by:

$$M = (1-x)M_4 + xM_3 = \chi(T) H \qquad (4)$$

where the behaviour of $\chi^{-1}(T)$ is a hyperbola,

$$\chi^{-1} = (T - \Theta)/C - \zeta/(T - \Theta') \qquad (5)$$

which, at high temperatures, approaches asymptotically a CW behaviour where C is given by equation 1 and

$$\Theta = [(1-x)^2 C_4^2 (\gamma_{44}+\gamma'_{44}) + 2x(1-x)C_3 C_4 (\gamma_{43}+\gamma'_{43}) + x^2 C_3^2 (\gamma_{33}+\gamma'_{33})] / C \qquad (6)$$

Using equation 5 for the fitting of the experimental data in the PM region, three independent parameters can be determined: $(\gamma_{44}+\gamma'_{44})$, $(\gamma_{33}+\gamma'_{33})$, and $(\gamma_{43}+\gamma'_{43})$, since first and second neighbours contributions cannot be separated.

Good fits were possible for $x \leq 0.10$ as shown in figure 3(a). For $x = 0$, equation 5 corresponds to a CW law with $C \approx C_4 = 1.875$, as described before. From the experimental value for $\Theta$ and equation 6 we derived $\Theta/C_4 = (\gamma_{44}+\gamma'_{44}) = -204$ mol/emu. In the same way, the susceptibility of the $x = 1$ sample would allow us to estimate $(\gamma_{33}+\gamma'_{33})$. The measurement in the G-type AFM $YMnO_3$ [10] gives $(\gamma_{33}+\gamma'_{33}) = -16$ mol/emu. However, the $YMnO_3$ compound is a much distorted perovskite and the tilting and rotation angles, related to the magnitude of the exchange interactions, are much larger than those measured in our case. For the less distorted $LaMnO_3$, a FM value $(\gamma_{33}+\gamma'_{33}) = +73$ mol/emu is obtained [20] for the high T pseudo-cubic phase. Thus, there is a large uncertainty associated with the probable magnitude of the $Mn^{3+}$-$Mn^{3+}$ interaction in our samples. We fitted the data for $0 < x \leq 0.10$ keeping only $(\gamma_{43}+\gamma'_{43})$ as an adjustable parameter and using a fixed value for $(\gamma_{33}+\gamma'_{33})$. We found that the results are rather insensitive to the value assumed for the $Mn^{3+}$-$Mn^{3+}$ interaction if we take $(\gamma_{33}+\gamma'_{33})$ in the range -16 mol/emu / +73 mol/emu, due to the relatively low concentration of $Mn^{3+}$ ions in the studied samples. The fitted curves, shown in figure 3(a), correspond to $(\gamma_{43}+\gamma'_{43}) = +361, +386$, and $+310$ mol/emu for $x = 0.05, 0.07$, and $0.1$, respectively. The $(\gamma_{43}+\gamma'_{43})$ parameter is always positive in these cases and larger than $(\gamma_{44}+\gamma'_{44})$, denoting a strong FM coupling for the $Mn^{4+}$-$Mn^{3+}$ pairs.

For $x \geq 0.15$, the susceptibility follows a CW law with a FM like $\Theta$, without the convex curvature observed for $x \leq 0.10$. It should be noticed that, although for $x \leq 0.10$ we obtained good agreement using the average C given by equation 1, the experimental results for $x \geq 0.15$ indicate progressively larger values for C (see figure 4). This observation may indicate a limit for the use of the ionic picture with Heisenberg interactions implicit in equations 2 and 3. If, for instance, the doping electrons were fully delocalized a better description might be obtained by assuming a perfect lattice of $Mn^{4+}$ ions magnetically coupled to the itinerant electrons, which, in turn, could be assumed to present a temperature independent susceptibility, $\chi_e(x) = x\chi_e^0$. As discussed in detail in [21] the susceptibility of the coupled system is given in this case by a *pure* CW law. The corresponding Curie constant is enhanced by the FM coupling between $Mn^{4+}$ magnetic moments and itinerant electrons. The enhancement is expected [21] to be proportional to x, as observed in the experiments (see figure 4).

In order to analyze the magnetic ordering temperature, $T_{mo}$, we must modify equations 2 and 3. Taking into account that the ordered state for $x = 0$ is a G-type AFM we divide the system into two interpenetrated sublattices, *a* and *b*, where $Mn^{4+}$ and $Mn^{3+}$ are randomly distributed with concentrations (1-x) and x, respectively. Although the ordered phase has not been determined for $x > 0$, neutron diffraction results [22] on $Ca_{1-x}Bi_xMnO_3$ indicate that the G-type AFM is preserved in this case up to $x \approx 0.1$. For higher values of x, different magnetic phases have been observed [23] in electron doped $Ca_{1-x}Sm_xMnO_3$. In all cases, the appearance of other magnetic phases was associated with structural transitions. As in our case the same orthorhombic structure was observed along the series, we have assumed valid the same division of the magnetic system into two sublattices. The modified equations 2 and 3 are:

$$M_4^{a,b}T = C_4[H + \gamma_{44}(1-x)M_4^{b,a} + \gamma'_{44}(1-x)M_4^{a,b} + \gamma_{43}xM_3^{b,a} + \gamma'_{43}xM_3^{a,b}] \quad (7)$$

$$M_3^{a,b}T = C_3[H + \gamma_{33}(1-x)M_3^{b,a} + \gamma'_{33}(1-x)M_3^{a,b} + \gamma_{43}xM_4^{b,a} + \gamma'_{43}xM_4^{a,b}] \quad (8)$$

Solving these equations for $H = 0$, the highest temperature that allows non trivial solutions for $M_i$ corresponds to the magnetic transition temperature $T_{mo}$. For $x = 0$, AFM ordering is achieved for $T_N = C_4(-\gamma_{44} + \gamma'_{44})$ and the values $\gamma_{44} = -134$ mol/emu and $\gamma'_{44} = -71$ mol/emu are derived from the measured $\Theta$ and $T_N = 123$ K. Notice that the second neighbours interaction is also AFM and almost 50% of the first neighbour interaction. For $x > 0$, two solutions are possible for $T > 0$ and are given by

$$T_{mo}^{(1)} \approx T_N(1-x) + (C_3C_4/T_N)(\gamma_{43} - \gamma'_{43})^2 x(1-x) \quad (9)$$

$$T_{mo}^{(2)} \approx [C_3C_4(\gamma_{43} + \gamma'_{43})^2 / \Theta_0] x \quad (10)$$

where $\Theta_0 = |\Theta|$ for $x = 0$.

For small x, $T_{mo}^{(1)} > T_{mo}^{(2)}$ and the ordering temperature is given by equation 9. The first term in this equation reflects the dilution of the $Mn^{4+}$ lattice and the second the effect of $Mn^{3+}$-$Mn^{4+}$ interactions, which always tend to increase $T_{mo}$. For this solution the magnetic state preserves the AFM alignment between $M_4^a$ and $M_4^b$. The $Mn^{3+}$ ions orient themselves FM with respect to their $Mn^{4+}$ first (or second) neighbours, provided that $\gamma_{43} > \gamma'_{43}$ ($\gamma_{43} < \gamma'_{43}$). Our experimental results show that the initial effect of doping is a decrease of $T_{mo}$, indicating that the weakening of the AFM interaction is dominant. As x increases, $T_{mo}^{(2)}$ is expected to become larger that $T_{mo}^{(1)}$ and $T_{mo}$ is given by equation 10.

Thus, $T_{mo}$ should present a minimum at a crossover concentration $x_c$. The predicted magnetic order for $x > x_c$ is still mainly AFM but the sublattice magnetizations are unbalanced. The experimental $T_{mo}(x)$ shows a minimum at about $x_c = 0.05$. According to equations 9 and 10 this crossover concentration corresponds to $\gamma_{43} \approx \gamma'_{43}$ and the values calculated for $T_{mo}$ in this case are shown in figure 5(b).

### 4. Montecarlo simulations

From our mean field analysis we may conclude that in the PM regime there are two differentiated regions as a function of the doping concentration x. For low x, the observed ferrimagnetic-like susceptibility is reasonable well described with an ionic model. For larger levels of doping a model assuming a fully delocalized $e_g$ electron is more appropriate for describing the experimental FM-like susceptibility.

In spite of the reasonable agreement regarding the qualitative behaviour of $\chi^{-1}(T)$ vs. T we should notice a couple of issues that deserve further attention:
i) The initial slope of $T_{mo}(x)$ vs. x is quite larger than the effect expected on the basis of $Mn^{4+}$ dilution alone, even if the always positive second term in equation 9 completely disappears by assuming $\gamma_{43} = \gamma'_{43}$.
ii) The observed increase of $T_{mo}$ for $x > x_c$ is much smaller that predicted by equation 10.
iii) The CW temperatures $\Theta(x)$ calculated using equation 6 (shown in figure 5(a)) indicate a high temperature behaviour that is mainly AFM-like. These values correspond to an asymptotic regime not reached in our experiments. The experimental values of $\Theta(x)$ reported in figure 5(a), obtained at finite temperatures, suggest an evolution of the system towards a FM regime more rapid than indicated by equation 6.

At this point we should take into consideration that the double exchange interaction may not be well represented in terms of the simple two site $Mn^{3+}$-$Mn^{4+}$ Heisenberg interaction implicit in our MF model for $x \leq 0.10$. This is so because DE is related to itinerant electrons that may occupy larger clusters or even be completely delocalized. In order to explore the degree of delocalization of the $e_g$ electrons under the competition of DE and SE interactions, we have performed Montecarlo simulations on an extended FM Kondo Hamiltonian that includes antiferromagnetic Heisenberg interactions for classical spins, as described in [24].

Our Montecarlo simulations show that the electrons doped in $Ca_{1-x}Y_xMnO_3$ are completely delocalized at high temperatures. Therefore, the material is homogeneous, all Mn ions present an intermediate valence state, and the AFM superexchange interactions are weakened by the same amount at each atomic site. This effect is similar to that discussed by Oles and Feiner [25] for the magnetic ordered phase. In the dilute limit the AFM weakening is proportional to the hoping matrix element [24] times the number of electrons per site, x in our case. The MC simulations give $\Theta(x)$ as shown in figure 5(a). The initial weakening of the AFM superexchange is followed, for $x > 0.05$, by a FM-like CW temperature ($\Theta > 0$) reproducing the experimental behaviour.

The ordering temperature was also derived from MC simulations and we found an important decrease for low doping, as observed in the experiments. As shown in figure 5(b) $T_{mo}$ begins to increase when the net interaction becomes FM ($\Theta(x) > 0$). Another interesting result of the Montecarlo simulations is that, in the paramagnetic phase, the doped electrons begin to localize and form mobile magnetic polarons of large spin when $T_{mo}$ is approached. Within this picture, the localizing process produces the observed convex curvature of $\chi^{-1}(T)$. Then our analysis of the magnetic susceptibility at temperatures near $T_{mo}$ points towards an inhomogeneous picture of the paramagnetic phase with the formation of magnetic polarons. We expect that these polarons maintain their structure and become well localized in the ordered phase ($T < T_{mo}$) embedded in the G-type AFM background. The magnetic moments of these polarons can be easily aligned along H even under low magnetic fields. This picture provides an explanation for the H dependence of the magnetization found at low temperatures [24]. Results derived from a single impurity model by Chen and Allen [4] that includes magnetic and elastic interactions agree with this picture. These and our MC results point towards an inhomogeneous picture for the low concentration samples rather than the homogeneous model proposed by van de Brink and Khomskii [26]. Neutron diffraction experiments could solve this matter.

## 5. Summary

Summarizing our results, we show in Figure 6 a schematic phase diagram as a function of the electron concentration x. Even if there are not conclusive experimental evidences for the transition between fully localized $e_g$ electrons and mobile polarons at a fixed x, it is interesting to notice in figure 3(a) a tendency of the experimental curves to deviate from the fitting curves towards a more FM behaviour at the highest temperatures. In the same way, for x = 0.15 (figure 3(c)) a noticeable change of slope is observed at $T \approx$ 300 K.

**Acknowledgement**

We acknowledge financial support from ANPCyT (Argentina), trough PICT 03-05266, and CONICET (Argentina) trough the H. A. PhD-fellowship.

**References**
[1] Zener C 1951 Phys. Rev. **81** 440.
[2] Wollan W O and Koehler W C 1955 Phys. Rev. **100** 545.
[3] Neumeier J J and Cohn J L 2000 Phys. Rev. B **61** 14319.
[4] Chen Y-R and Allen P B 2001 Phys. Rev. B **64** 064401.
[5] Granado E, Moreno N O, Martinho H, García A, Sanjurjo J A, Torriani I, Rettori C, Neumeier J J and Oseroff S B 2001 Phys. Rev. Lett. **86** 5385.
[6] Neumeier J J and Goodwin D H, 1999 J. Appl. Phys. **83** 5591.
[7] Aliaga H, Causa M T, Alascio B, Salva H, Tovar M, Vega D, Polla G, Leyva G and König P 2001 J. Magn. Magn. Mater. **226-230** 791.
[8] Vega D, Polla G, Leyva A. G, König P, Lanza H, Esteban A, Aliaga H, Causa M T, Tovar M and Alascio B, 2001 J. Solid State Chem. **156** 458.
[9] Knížek K, Jirák Z, Pollert E and F. Zunová 1992 J. Solid State Chem. **100** 292. . Pollert E, Krupicka S and Kuzmicová E 1982 J. Phys. Chem. Solids **43** 1137.
[10] Agüero O, Leyva A G, König P, Vega D, Polla G, Aliaga H and Causa M T 2002 Physica B **320** 47.
[11] Rettori C private communication. Radaelli P G, Cox D E, Marezio M, Cheong S-W, Schiffer P E and Ramírez A P (1995) Phys. Rev. Lett. **75** 4488. Subías G, García J, Blasco J Proietti M G (1998) Phys. Rev. B **57** 748.
[12] Jirák Z, Krupicka S, Šimša Z, Dlouhá M and Vratislav S (1985) J. Magn. and Magn. Mater. **53** 153.
[13] Blaso J, Riter C, García J, de Teresa J M, Pérez-Cacho J and Ibarra M R (2000) Phys. Rev. **62** 5609.
[14] Vega D, Ramos C, Aliaga H, Causa M T, Alascio B, Tovar M, Polla G, Leyva G, König P and Torriani I 2002 Physica B **320** 37.
[15] Briático J, Allub R, Alascio B, Caneiro A, Causa M T and Tovar M 1996 Phys. Rev. B **53** 14020.
[16] Lines M E 1967 Phys. Rev. **164** 736.
[17] Müller K A 1959 Phys. Rev. Lett. **2** 341.
[18] Causa M T, Tovar M, Caneiro A, Prado F, Ibáñez G, Ramos C A, Butera A, Alascio B, Obradors X, Piñol S, Tokura Y and Oseroff S B 1998 Phys. Rev. B **58** 3233
[19] Huber D L, Alejandro A, Caneiro A, Causa M T, Prado F, Tovar M and Oseroff S B 1999 Phys. Rev. B **60** 12155.
[20] Tovar M, Alejandro G, Butera A, Caneiro A, Causa M T, Prado F, and Sánchez R 1999 Phys. Rev. B **60** 10199.
[21] Tovar M, Causa M T, Butera A, Navarro J, Martínez B, Fontcuberta J and Passeggi M. C. G Cond-mat/0205187 and 2002 Phys. Rev. B **66** 024409.
[22] Santosh P N, Goldberg J, and Woodward P M 2000 Phys. Rev. B **62** 14928.
[23] Mahendiran R, Maignan A, Martin C, Hervieu M, and Raveau B 2000 Phys. Rev. B **62**, 11644.
[24] Aliaga H, Causa M T, Tovar M and Alascio B 2002 Physica B **320**, 75.
[25] Oles A M and Feiner L F 2002 Phys. Rev. B **65** 052414.
[26] van den Brink J and Khomskii D 1999 Phys. Rev. Lett. **82** 1016.

**Figure captions**

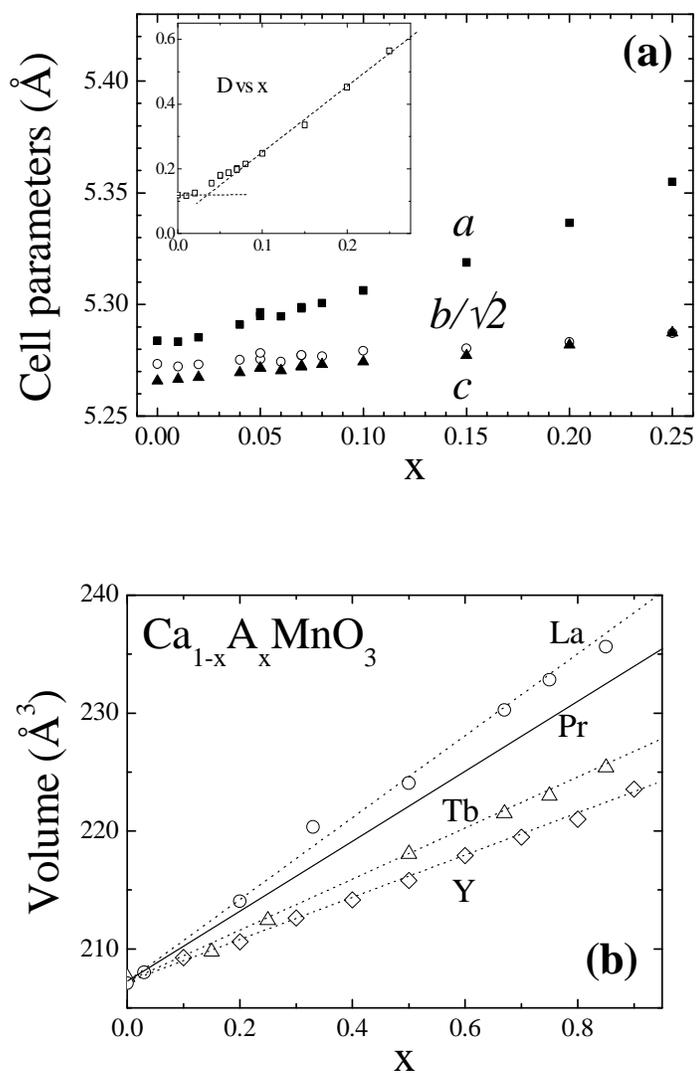

Figure 1.
(a) Variation of the cell parameters with the Y concentration, x.  In the inset, the orthorhombic distortion D vs. x is shown.  (b) Cell volume vs. x for $Ca_{1-x}A_xMnO_3$ with A = La, Pr, Tb and Y.

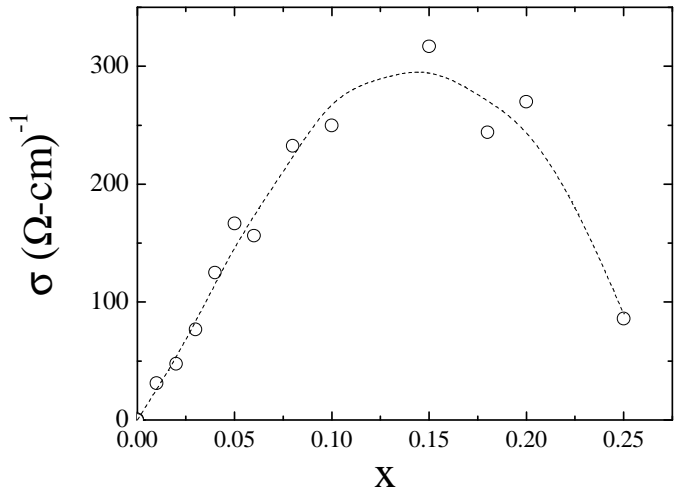

Figure 2.
Electrical conductivity for different values of x measured at room temperature.

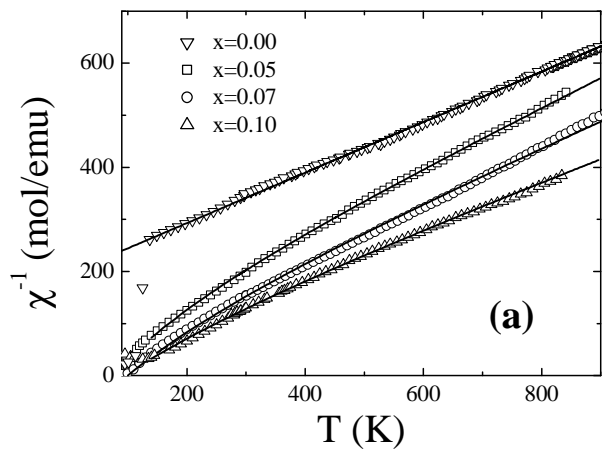

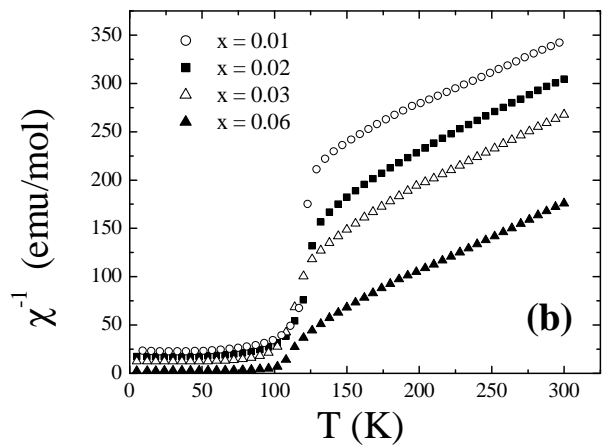

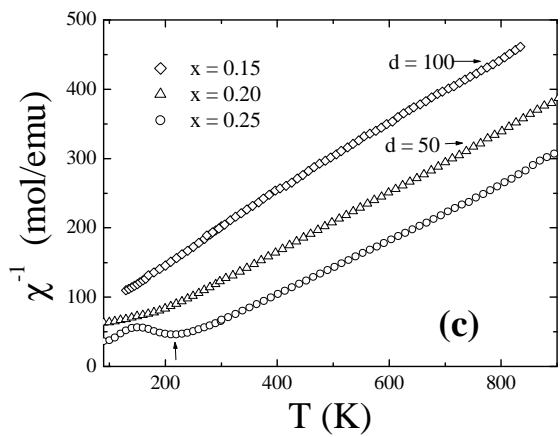

Figure 3.
$\chi^{-1}(T)$ vs. T for different values of x. In (a), the lines are the best fits to equations 2 and 3. In (b) a detail of the behaviour near the magnetic transition is shown. The curves for x = 0.15 and 0.20, in (c), were displaced vertically a quantity d, as indicated. Arrow: CO transition for x = 0.25.

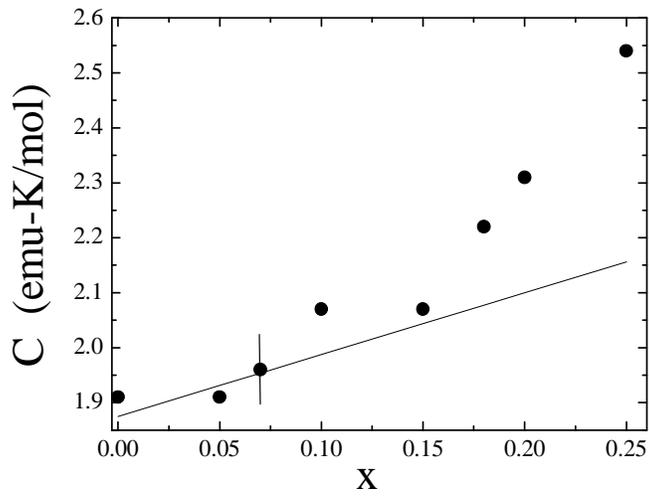

Figure 4
Curie constant C(x) vs. x. The line is C(x) calculated with equation 1.

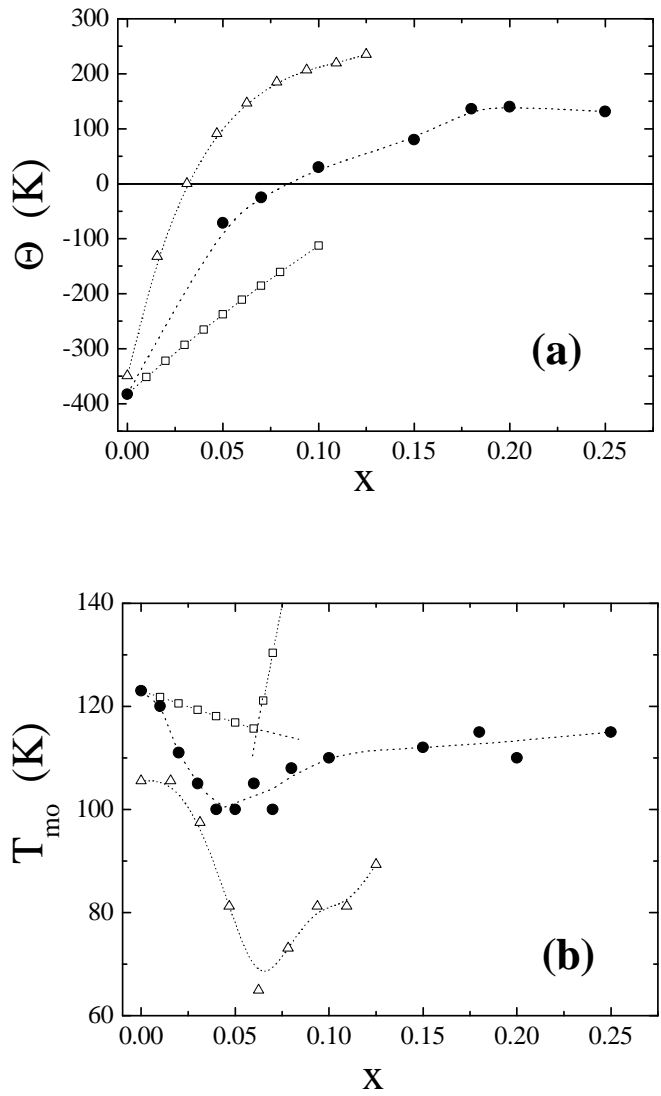

Figure 5
Experimental values (solid circles) for: (a) CW temperature Θ(x) and (b) magnetic ordering temperature $T_{mo}(x)$. Squares and triangles are, respectively, mean-field and Montecarlo calculations. Lines are eye guides.

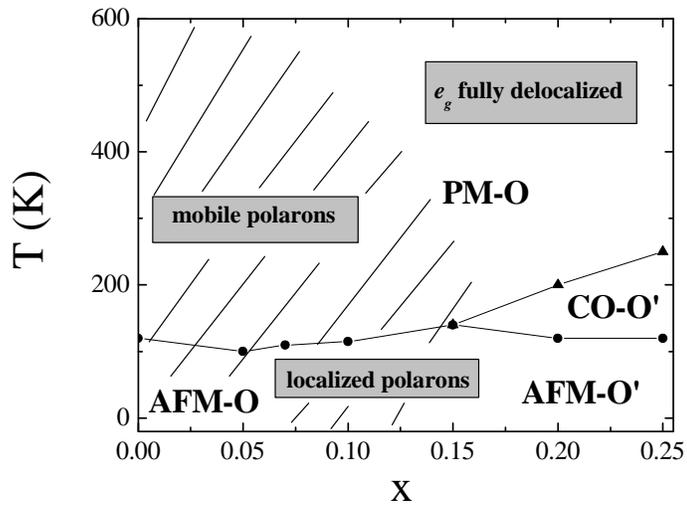

Figure 6
Schematic phase diagram T vs. x for $Ca_{1-x}Y_xMnO_3$. The magnetic and charge ordering transition temperatures are shown by circles and triangles, respectively. At high temperatures, in the PM region, the orthorhombic phase O is observed for all x. At the lowest temperatures, a transition from O to O' structures takes place at $x \approx 0.15$. The region marked with lines indicates the range where magnetic polarons are expected to be present.